\documentclass[conference]{IEEEtran}
\IEEEoverridecommandlockouts

\usepackage{tikz}
\usepackage{graphics}
\usepackage{lscape}
\usepackage{adjustbox}
\usepackage{tabularx}
\usepackage{float}

\usepackage{listings}
\usepackage{color}
\usepackage{commath}
\usepackage{mathtools}
\usepackage{acronym}
\usepackage[toc, acronym]{glossaries}
\usepackage{enumitem}
\usepackage{wrapfig}

\usepackage{hyperref}
\hypersetup{
    colorlinks=true,
    linkcolor=blue,
    filecolor=magenta,      
    urlcolor=cyan,
    pdftitle={Overleaf Example},
    pdfpagemode=FullScreen,
    }

\usepackage{comment}

\usepackage{pgfplots}

\usepackage[utf8]{inputenc}
\usepackage{csquotes}

\usepackage{dirtytalk}
\definecolor{Cyan}{rgb}{0.58,1,1}
\definecolor{LightCyan}{rgb}{0.88,1,1}
\definecolor{Purple}{rgb}{0.50,0,0.50}
\definecolor{Gray}{gray}{0.85}
\definecolor{coolblack}{rgb}{0.0, 0.18, 0.39}
\definecolor{copper}{rgb}{0.72, 0.45, 0.2}
\definecolor{goldenpoppy}{rgb}{0.99, 0.76, 0.0}
\definecolor{mahogany}{rgb}{0.75, 0.25, 0.0}
\usepackage{url}
\usepackage{multirow}
\usepackage{xfrac}
\usepackage{colortbl}
\usetikzlibrary{patterns}
\usepackage{graphicx}
\usepackage{subcaption}

\usepackage[bottom]{footmisc}

\usepackage{cite}
\usepackage{amsmath,amssymb,amsfonts}

\usepackage{algorithmic}
\usepackage{graphicx}
\usepackage{pgfplots}
\usepackage{enumitem}
\usepackage[utf8]{inputenc}
\usepackage{dirtytalk}
\usepackage{tabularx}
\usepackage{tikz}
\usepackage{textcomp}
\usepackage{pgfplots}
\pgfplotsset{compat=1.12}
\usetikzlibrary{pgfplots.groupplots}
\usetikzlibrary{plotmarks}
\usetikzlibrary{matrix}
\usepgfplotslibrary{groupplots}
\pgfplotsset{compat=newest}
\pgfplotsset{
    every non boxed x axis/.style={} 
}

\usepackage[absolute,overlay]{textpos}

\begin{document}
\title {CxSE: Chest X-ray Slow Encoding CNN for COVID-19 Diagnosis}

\author{{Thangarajah Akilan}

\thanks{Thangarajah Akilan is with the Department of Software Engineering, Lakehead University, Thunder Bay, ON, Canada. (e-mail: \{takilan\}@lakeheadu.ca).}

}

\maketitle

\begin{abstract}
The coronavirus continues to disrupt our everyday lives as it spreads at an exponential rate.  It needs to be detected quickly in order to quarantine positive patients so as to avoid further spread. This work proposes new convolutional neural network (CNN) architecture called 'slow Encoding CNN. The proposed model's best performance wrt Sensitivity, Positive Predictive Value (PPV) found to be  SP=0.67, PP=0.98, SN=0.96, and PN=0.52 on AI AGAINST COVID19 - Screening X-ray images for COVID-19 Infections competition's test data samples. SP and PP stand for the Sensitivity and PPV of the COVID-19 positive class, while PN and SN stand for the Sensitivity and PPV of the COVID-19 negative class.
\end{abstract}



\section{Introduction}
Coronavirus, also known as COVID-19 was discovered in Wuhan, China, in December of 2019~\cite{who}. COVID-19 has many strains and can infect animals and humans. COVID-19 is hard to detect because it has common symptoms such as cold and flu. The symptoms also range in seriousness depending on the person's immune system. Symptoms can take up to 14 days to appear after exposure. Because of this, the public disregards them as everyday common flu or cold. COVID-19 is spread through respiratory droplets when you cough, sneeze, and touch~\cite{govca}. COVID-19 spread has become so severe it is shutting down our economies. There are over 127 million worldwide cases and over 2.7 million deaths as of March 29, 2021 and rising daily \cite{[2]asean}. Chest X-rays and CT scans can be conducted quickly and efficiently for detecting COVID-19~\cite{Wang2020}. It will allow the radiologist, pathologist, and physicians to properly learn the condition of the COVID-19 affected patients for additional care and drive specific clinical solutions to safe lives. The quicker the detection, the quicker the patient will receive treatment and can be put in quarantine to avoid further spread.

This work introduces a new CNN-based solution with slow feature learning strategy to predict if the person is affected with COVID-19 using the patient's chest x-ray. The model is solely trained and tested on COVIDx CXR-2 dataset provided by the \textit{AI against COVID-19: Screening X-ray Images for COVID-19 Infections} competition, which is available at 
\href{https://www.kaggle.com/andyczhao/covidx-cxr2}{data set}. It comprises of over 16000 ($480 \times 480$) chest X-ray scans taken from 15000 patients from across the world (minimum of 51 participating countries). We follow the exact training and test sets splits containing positive and negative classes to indicate COVID or non-COVID cases, and evaluate the proposed model on the hold out set that do not provide the ground truths. For performance computation we submit our model's binary label predictions to the evaluating site \href{https://eval.ai/web/challenges/challenge-page/925/leaderboard}{leaderboard} and receive the results.

\section{Literature Review}\label{Section II}

To tackle the surge of COVID-19, the AI community has been actively developing efficient solutions for automatic detection of COVID-19 patients  from various sources as an alternative/supportive to the conventional time consuming diagnosis procedures. Among them the vision-based (e.g., the medical images, like X-rays, CT scans) classification models show promising results. 


As data scarcity is a longstanding issue in the medical machine learning, most of the researchers go by transfer learning (TL) approach. In this direction, Gozes~\textit{et al.}~\cite{Gozes2020} focus using the well-known UNet~\cite{ronneberger2015u} and ResNet50-2D~\cite{he2016deep} architectures for CT scan-based COVID-19 patient classification, quantification and tracking for the patients. 
Similarly, Wang \textit{et al.}~\cite{[5]Wang2020} use ResNet18 architecture, and  Ali \textit{et al.}~\cite{[6]narin2021automatic} employ ResNet50, InceptionV3 and Inception-ResNetV2~\cite{szegedy2017inception} models. 
The main aim of these TL techniques is to extract features from the small number of medical images leveraging an exhaustively trained CNN on large-scale data, then a shallow classifiers, like decision tree and SVM on the extracted feature sets. Relatively, such approaches work well. That being said, they are over dependant on the pretrained backbone models.  




In response to that, there are few attempts where researchers carefully \cite{Wang2020}, \cite{aboutalebi2021covid} architect new deep learning models specifically for the detection COVID-19. In this line, this work CNN model is trained from scratch on Chest X-rays to classify COVID-19 cases.


\section{Proposed Slow Encoding CNN}

\begin{figure}[!tb]
\centering
\includegraphics[width=0.55\columnwidth, trim={3.0cm 2.5cm 2.5cm 2.5cm},clip]{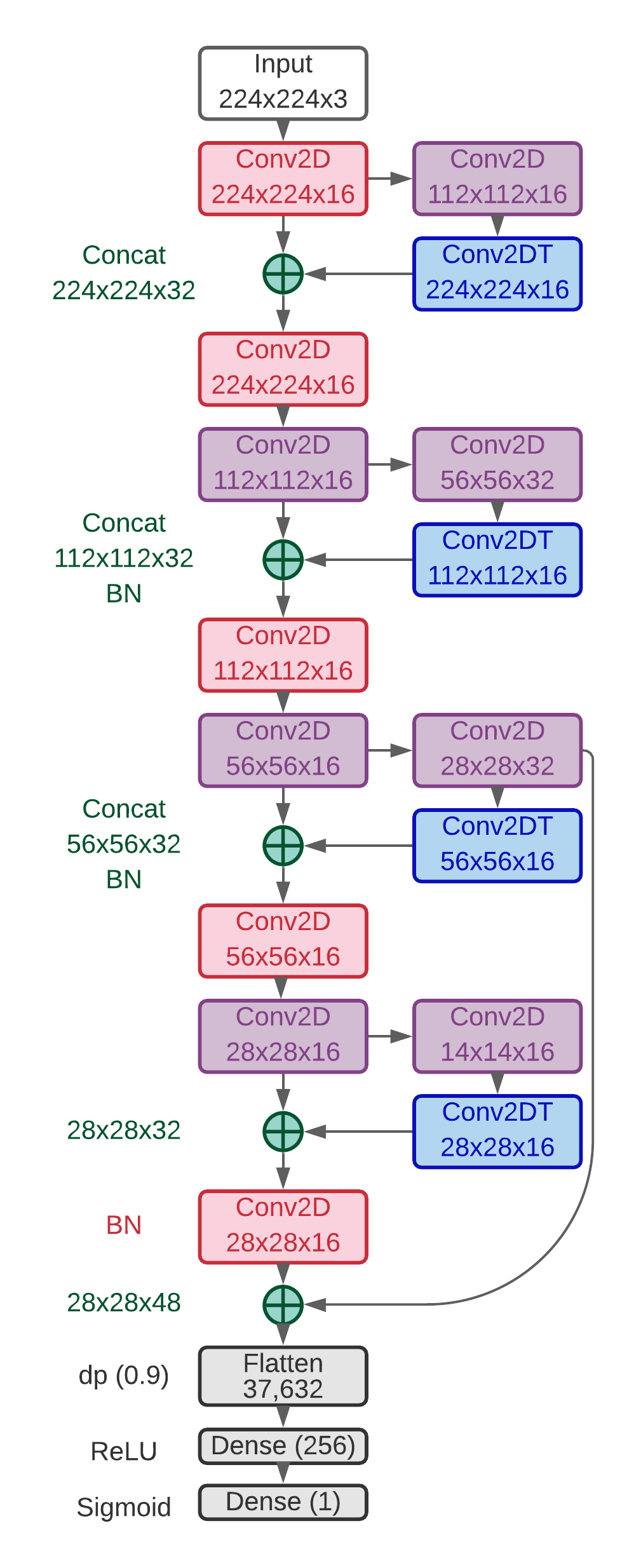} 
\caption{The proposed slow encoding Convent classifier for COVID-19 detection using chest X-rays. All Conv2D layers use ReLU activation. }
\label{fig:sendecmodel}
\end{figure}

Earlier in \cite{Akilan2020, Akilan2018tvt, akilan2018}, we introduced a novel encoder-decoder (EnDec) foreground segmentation architecture that perform feature learning twice at every stage of down-sampling processing. It has two subnets (cf.~Fig.~\ref{fig:sendecmodel}): encoder (spatial subsampling) and decoder subnet (up-sample the lower dimensional bottle neck output feature map of the encoder back to the original dimension of the image). In the encoder subnet, a spatial input ($layer_i$) is transformed through a spatial sub-sampling convolutional (Conv) operations (($layer_{i+1}$)), followed by an up-sampling operation using a transpose convolution (($layer_{i+2}$)) that generates exactly matching spatial feature maps to the previous layer’s input feature dimension. Now, the up-sampled feature maps are aggregated depth wise with the original input features ($layer_i$). Then, the aggregated new feature maps are encoded using a spatial subsampling Conv layer. In this way, an input feature map in sub-sampling stage is learnt twice before completely moving to the next-level of lower spatial dimension. Following that, this work upcycle the model by removing decoder subnet and refurbish it by adding dense layers on the top with a Sigmoid classifier targeting the COVID-19 patient classification using Chest X-ray images. Figure ~\ref{fig:sendecmodel} depicts the proposed slow encoding Convnet model. In total the model has 9,692,865 trainable parameters.

\section{Data-set}\label{Section III}











\section{Experimental Analysis}\label{Section VII}

\subsection{Training Time Analysis}

\subsubsection{Training History}

\begin{figure}[!htb]
\centering
\includegraphics[width=1.0\columnwidth, trim={0.25cm 0cm 0.25cm 0cm},clip]{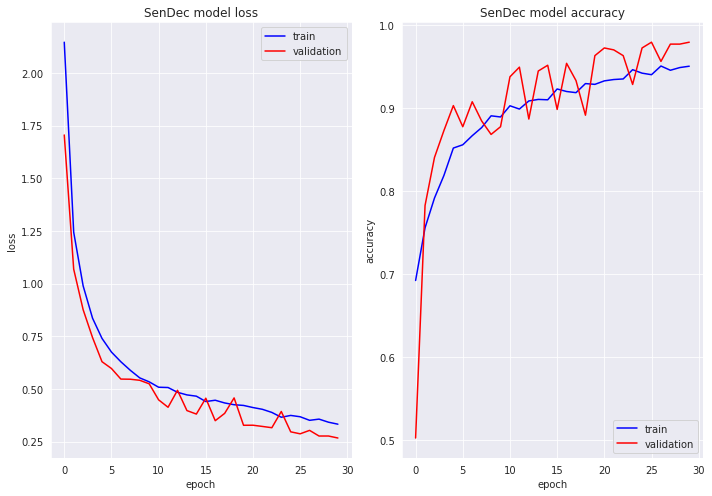} 
\caption{The training history of CxSE.}
\label{fig:traininghis}
\end{figure}

The proposed CxSE model was trained for 30 epochs on the given train set on the competition site using Adamax optimizer with learning rate of 0.001. The training history is show in Fig.~\ref{fig:traininghis}.

\subsubsection{Testing Results}

\begin{figure}[!htb]
\centering
\includegraphics[width=0.9\columnwidth, trim={0cm 0cm 0cm 0cm},clip]{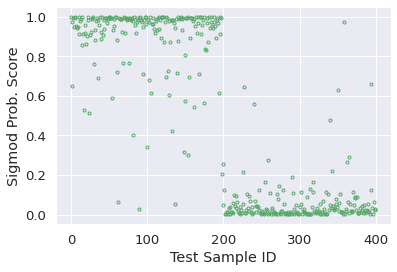} 
\caption{CxSE's sigmoid probability scores on test set. }
\label{fig:testsetscores}
\end{figure}

\begin{figure}[!htb]
\centering
\includegraphics[width=0.6\columnwidth, trim={0cm 0cm 0cm 0cm},clip]{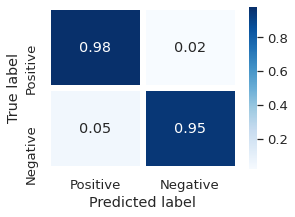} 
\caption{Proposed CxSE's performance on test set. }
\label{fig:confusionmat}
\end{figure}

\begin{figure}[!htb]
\centering
\includegraphics[width=0.9\columnwidth, trim={0cm 0cm 0cm 0cm},clip]{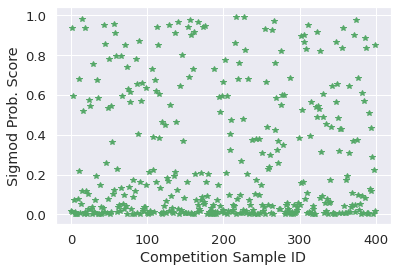} 
\caption{CxSE's sigmoid probability scores on competition set. }
\label{fig:competitionsetscore}
\end{figure}

The proposed model is tested on the given test set on the competition site. The probability scores produces (cf.~Fig.~\ref{fig:testsetscores}) by the model is converted into class labels using a threshold of 0.5 and it is compared against the ground truths. The model's performance is shown by a confusion matrix in Fig.~\ref{fig:confusionmat}. 
Similarly, the model's probability scores on the competition set (cf.~Fig.~\ref{fig:competitionsetscore}) is also converted using the same threshold and uploaded to the competition's \hyperlink{https://eval.ai/web/challenges/challenge-page/925/leaderboard/2424}{evaluation site} to get the model's overall performance as tabulated in Table~\ref{tab:leaderboardscore}. 

\begin{table}[!ht]
    \centering
    \begin{tabular}{|c|c|}
    \hline 
         \textbf{Metrics} &  \textbf{Score}\\ \hline \hline
        {SP} &  0.67\\\hline
        {PP} & 	0.98 \\\hline
        {SN} & 	0.96\\\hline
        {PN} & 	0.52 \\\hline
        {Overall points} & 	12.80\\ \hline
    \end{tabular}
    \caption{CxSE's COIVD-19 classification performance on competition set (cf. \href{{https://eval.ai/web/challenges/challenge-page/925/leaderboard/}}{leaderboard} - Participant team: MVLC)}
\label{tab:leaderboardscore}
\end{table}


\section{Conclusion}\label{Section VIII}

This work has introduced a new feature learning DCNN aiming for COVID-19 diagnosis using chest x-rays. This fist stage of our development shows promising outcome. The future work is dedicated to improving the model's learning ability through class-agnostic semi-supervised training approach.



\section*{Acknowledgements}\label{Acknowledgements}
This work acknowledges IEEE SIGHT MTL, VIP lab, and DarwinAI the organizers of ``AI Against COVID19 - Screening X-ray images for COVID-19 Infections''.

\bibliographystyle{IEEEtran}
\bibliography{main.bib}


\end{document}